\documentclass[aps,prl,preprint,groupedaddress]{revtex4}
\usepackage{amssymb}
\usepackage{amsmath}
\usepackage{graphicx}

\begin{document}

\title{Field-induced vortices in weakly anisotropic ferroelectrics}
\author{Ana\"{\i}s Sen\'{e}}
\affiliation{Laboratory of Condensed Matter Physics, University of Picardie Jules Verne,
Amiens, 80039, France \\
anais.sene@gmail.com}
\author{Laurent Baudry}
\affiliation{Institut d'Electronique, de Micro\'electronique et de Nanotechnologie (IEMN)
\\
UMR CNRS 8520, D\'epartement Hyperfr\'equences et Semi-conducteurs, \\
Universit\'e des Sciences et Technologies de Lille, \\
Avenue Poincar\'e, BP 69, 59652 Villeneuve d'Ascq Cedex, France\\
laurent.baudry@iemn.univ-lille1.fr }
\author{Igor A. Luk'yanchuk}
\affiliation{Laboratory of Condensed Matter Physics, University of Picardie Jules Verne,
Amiens, 80039, France \\
lukyanc@ferroix.net}
\author{Laurent Lahoche}
\affiliation{Roberval Laboratory, University of Technology, Compiegne, 60200, France\\
laurent.lahoche@gmail.com}
\date{\today }

\begin{abstract}
In the micro- and nanoscale ferroelectric samples, the formation and the
growth of domains are the usual stages of the polarization switching
mechanism. By assuming the weak polarization anisotropy and by solving the
Ginzburg-Landau-Khalatnikov equation we have explored an alternative
mechanism which consists in ferroelectric switching induced by vortex
formation. We have studied the polarization dynamics inside a ferroelectric
circular capacitor where switching leads to formation of a metastable vortex
state with a rotational motion of polarization. Our results are consistent
with recent first-principa simulations [I. I. Naumov and H. X. Fu, Phys.
Rev. Lett. 98, 077603 (2007)] and with experiments in PbZr$_{0.2}$Ti$_{0.8}$O%
$_{3}$ [A.Gruverman et al, J. Phys. Condens. Matter 20 342201(2008)] and
demonstrate that vortex induced polarization switching can be the effective
mechanism for circular nano-capacitors.
\end{abstract}

\maketitle


\bigskip

For a long time, lot of attention has been given to study of finite size
surface and interface effects in ferroelectric materials because of their
fundamental interest and potential applications in electronic devices. As an
example the up- and down- polarized domains can serve as the binary
information units in Ferroelectric Random Access Memories (FRAM) \cite%
{Scott}. In micro- and nanoscopic thin films the domains can form
the periodic thermodynamically stable structures, provided by
interplay of ferroelectric condensation energy and electrostatic
energy \cite{PRL}. Stability of ferroelectric domains and switching
conditions are therefore the crucial aspects from the viewpoint of
the reliability of ferroelectric devices.

When an electric field is applied the polarization changes its direction.
Two mechanisms of polarization reversal are possible: the discontinuous
overturn from \textquotedblleft up\textquotedblright\ to \textquotedblleft
down\textquotedblright\ of the polarization and the Bloch one when the
direction of polarization continuously changes conserving the modulus of
amplitude.

The objective of the present communication is to study the Bloch
polarization dynamics in ferroelectric capacitor and to discuss the
results, relevant for their application in ferroelectric devices.
Our static and dynamic results are consistent with first-principa
simulations of stable vortex states by Naumov et all \cite{Naumov}
and can explain the existence
of metastable vortex state experimentally observed by Gruverman et all \cite%
{Gruverman}.

\begin{figure}[h]
\includegraphics  [width=7cm]{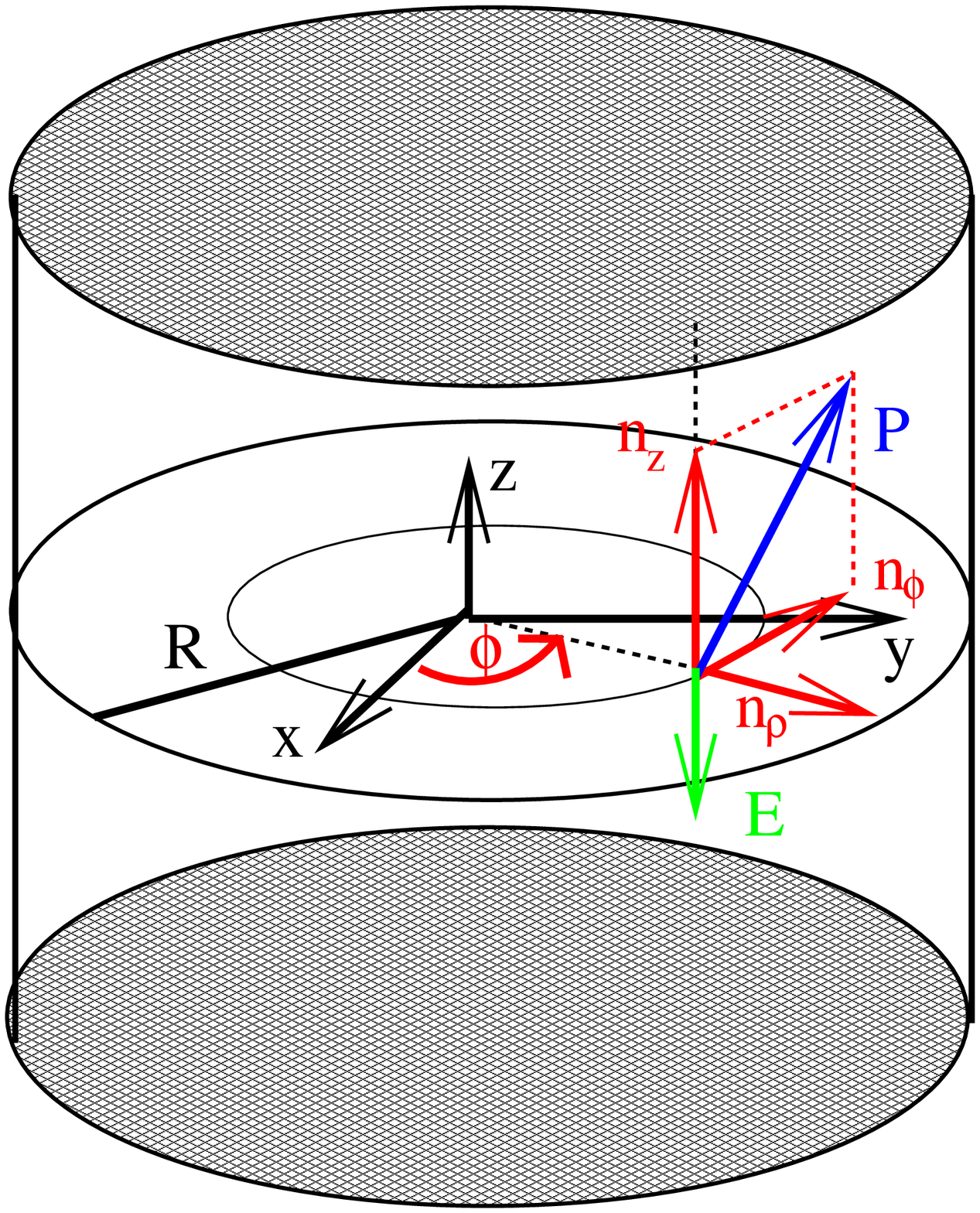}
\caption{Capacitor-like geometry of the model}
\label{repere}
\end{figure}

The problem set up having the geometry of the ferroelectric circular
capacitor of radius R shown in Fig \ref{repere}. The uniform external field E= - E $%
\mathbf{n}_{\mathbf{z}}$ is applied perpendicular to capacitor
plats. In addition no depolarizing charge is induced in the bulk
i.e. condition
\begin{equation}
\mathsf{div}\text{ }\mathbf{P}=\boldsymbol{0}  \label{div}
\end{equation}%
is satisfied. The surface depolarization field is screened by electrodes.

The general form of the texture of polarization satisfying the condition (%
\ref{div}) can be present in cylindrical coordinate P ($P_{\rho }$, $\
P_{\phi }$, $P_{z}$) as:

\begin{equation}
P_{\rho }=0,\ \ \mathbf{P}=\ P_{\phi }(\rho )\mathbf{n}_{\phi }+P_{z}(\rho )%
\mathbf{n}_{z}
\end{equation}%
We also assume that the polarization at the boundary of the capacitor is
fixed.
\begin{equation}
\mathbf{P}(\rho =0)=\mathbf{P}(\rho >R)=P_{z}\mathbf{n}_{\mathbf{z}}
\label{CL}
\end{equation}%

The Bloch switching mechanism implies that the only Goldstone mode is
involved into polarization reverse and the amplitude of polarization remains
constant.

The key feature of the Bloch switching process is the weak interaction of
polarization with the crystal axes that we believe takes place in PbZr$_{x}$%
Ti$_{1-x}$O$_{3}$ close to morphotropic point.

Neglecting for beginning the polarization anisotropy at all we present the
free energy F of the system as
\begin{equation}
F=K^{\prime }\bigtriangledown _{i}P_{i}\bigtriangledown _{j}P_{j}+K^{\prime
\prime }\bigtriangledown _{i}P_{j}\bigtriangledown _{i}P_{j}-\frac{\varkappa
}{4\pi }P_{i}E_{i}  \label{fonct}
\end{equation}

(tensor summation is used).

The first two terms are the gradient energy and the last one is interaction
between the polarization and the applied field. Coefficients K', K''
corresponds to the Ginzburg coefficients and $\varkappa $ to the
permittivity.

Numerical minimization of the free energy (\ref{fonct}) should give the
polarization distribution in the presence of an applied field. Fig. \ref%
{profil} presents the polarization profiles for different values of applied
field. Below some critical electrical field E$_{z}^{cr}$ the equilibrium
state corresponds to a uniform \textquotedblleft up\textquotedblright\
polarization state ($\ P_{\phi }$ =0, $P_{\mathbf{z}}$ =1) that means that
uniform polarization distribution is stable at E$\prec $ E$_{z}^{cr}$ . The
calculation of the numerical value of E$_{z}^{cr}$ will be given elsewhere.

Above the critical electrical field the polarization starts to
deviate from the uniform state and exhibits both non zero $\ P_{\phi
}$ component, which form the vortex structure \cite{Bellaiche}.

We observe that above the critical field, the equilibrium polarization
distribution is asymmetric. Moreover this asymmetry is enhanced by the
increase of the electrical field.

\begin{figure}[tbp]
\centering
\includegraphics  [width=16cm]{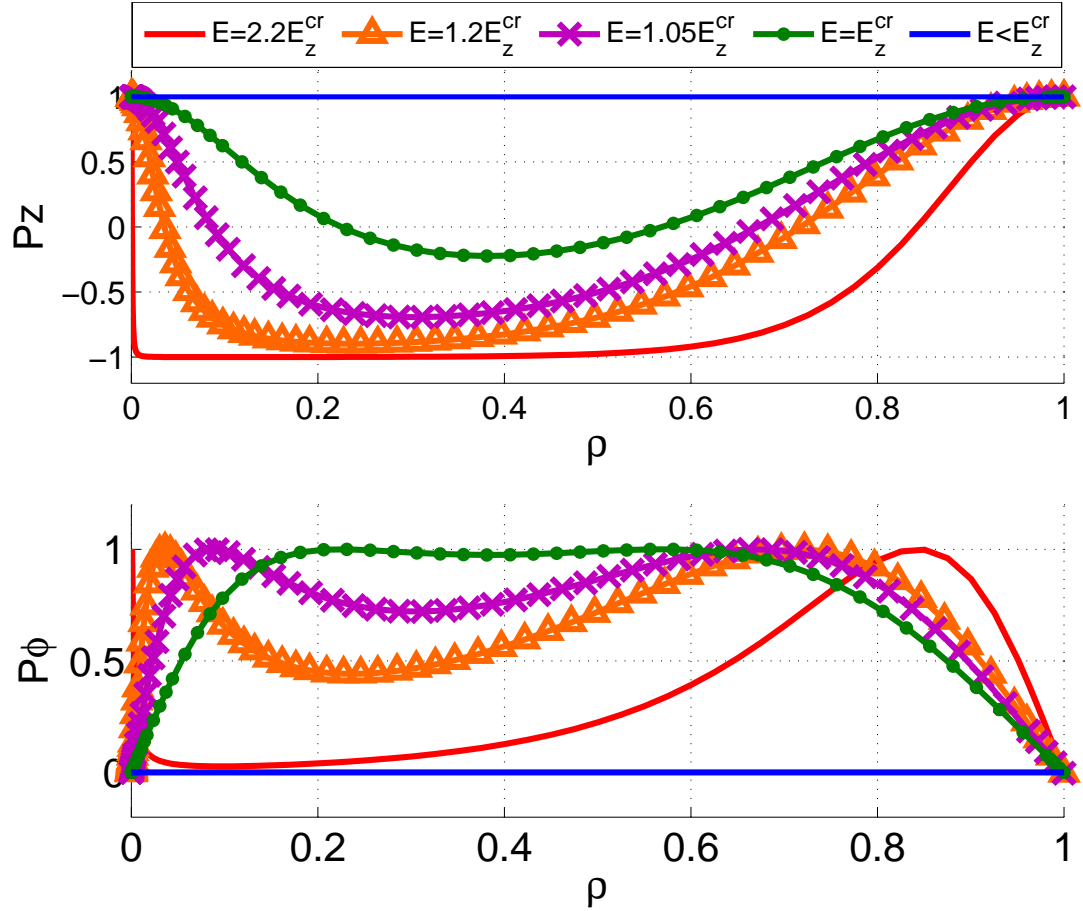}
\caption{Evolution of profile of polarization P ($\ P_{\phi }$,
$P_{z}$) as function of dimensionless radius for different values of
applied field} \label{profil}
\end{figure}

To better understand the formation of the vortex structure, we have studied
the evolution of the polarization from the initial homogeneous up state to
the final vortex structure as a function of time. The dynamic of the system
can be described by the Ginzburg-Landau-Khalatnikov equation \cite{Cottam}%
\begin{equation}
\frac{\partial P_{i}}{\partial t}=-\gamma _{i}\frac{\delta F}{\delta P_{i}}
\label{LKeq}
\end{equation}%
where $\gamma _{i}$ are kinetic coefficients and i =\{$\phi ,z$\}

\begin{figure}[tbp]
\centering
\includegraphics [width=18cm] {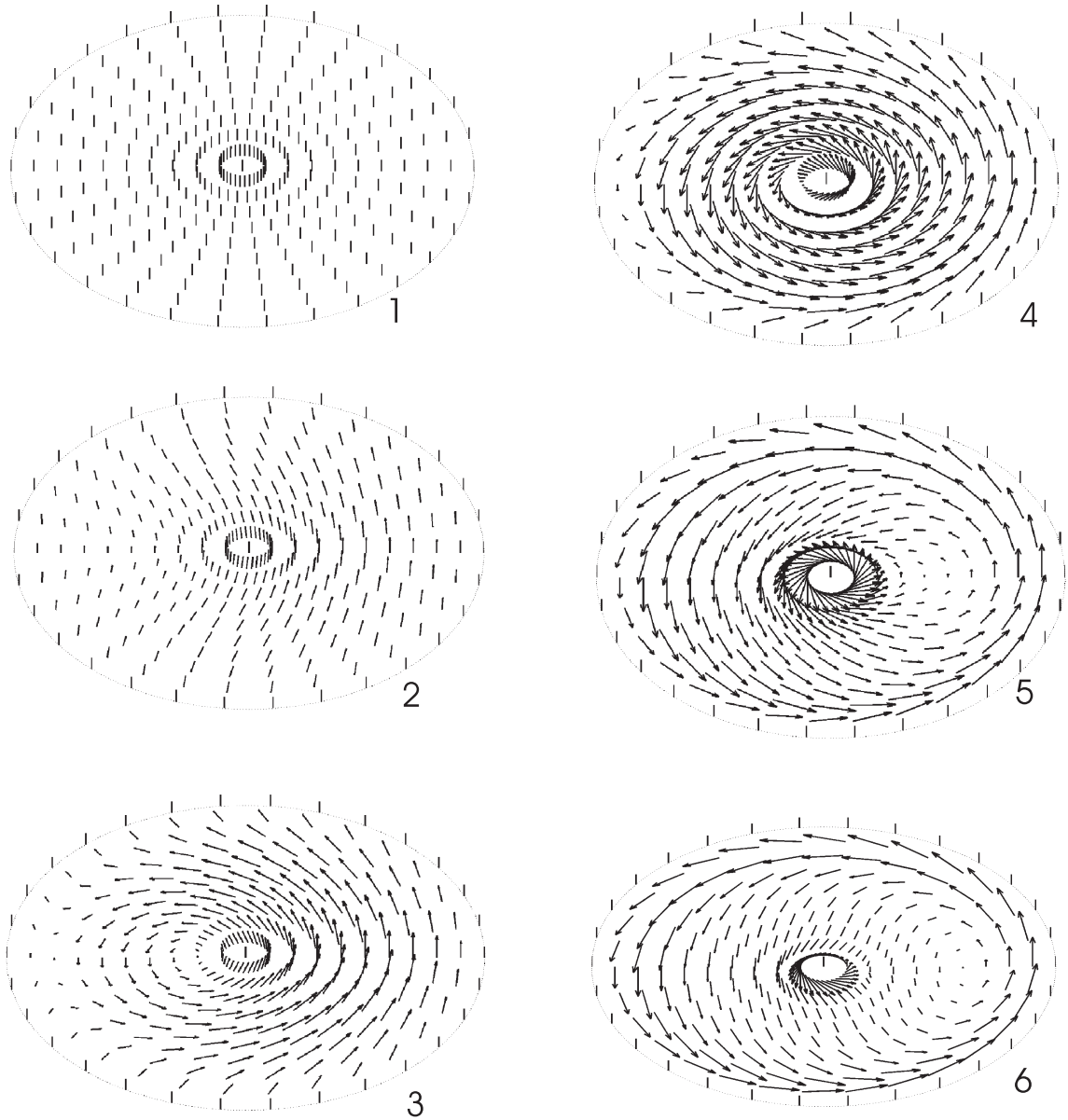}
\caption{Time evolution of polarization pattern between initial up
polarized state and final doughnut-shape state} \label{dynamiq}
\end{figure}

Dynamical behavior of polarization pattern is presented in Fig. \ref{dynamiq}%
. We observe the time evolution of the polarization vector from an initial
polarized state \textquotedblleft up\textquotedblright . It progressively
rotates and in the intermediate state the polarization component $P_{z}$
changes in sign. The final state exhibits a doughnut shape i.e. with
\textquotedblleft up\textquotedblright\ polarization state at the center and
at the electrode perimeter limit, and with \textquotedblleft
down\textquotedblright\ polarization state in the interior region.

Up to now we considered the particular case of the uniform gradient energy
functional (\ref{fonct}). We believe however that square-like anisotropy
terms can be considered as perturbation that will quadratically deform the
vortex shape but not change the principal conclusions about the switching
dynamics.

In conclusion, using thermodynamical approach and assuming the weak
polarization anisotropy and the absence of depolarizing field we
have determined the equilibrium state in the presence of an applied
field for ferroelectric system with circular electrodes. We obtained
the unusual domain pattern which looks like a doughnut and that were
previously observed experimentally \cite{Gruverman}. Time dependent
simulation allowed us to describe the vortex formation.

The part of the work was done in the frame of FP7-IRSES-Robocon program.

\end{document}